\PassOptionsToPackage{hyphens}{url}
\documentclass[submission,copyright,creativecommons]{eptcs}
\hypersetup{breaklinks=true}

\providecommand{\event}{ACL2-2025}

\usepackage[all]{hypcap} 
\usepackage{alltt}
\usepackage{amssymb}
\usepackage{enumitem}
\usepackage[T1]{fontenc}
\usepackage{tikz}
\usepackage{underscore}
\usepackage{xcolor}

\newcommand{\secref}[1]{Section \ref{#1}}
\newcommand{\figref}[1]{Figure \ref{#1}}

\newcommand{\code}[1]{\texttt{#1}} 
\definecolor{commentgray}{gray}{0.4}
\newenvironment{bcode} 
 {\begin{quote}\small\begin{alltt}}
 {\end{alltt}\end{quote}}

\newcommand{\citecode}[2]
  {\cite[\href{https://github.com/acl2/acl2/tree/master/#1}{\texttt{#2}}]
        {acl2-code}}
\newcommand{\citeman}[2]
  {\cite[\href{http://acl2.org/manual?topic=#1}{\texttt{#2}}]
        {acl2-manual}}

\begin{document}


\title{A Formalization of the Yul Language and \\
       Some Verified Yul Code Transformations}

\author{Alessandro Coglio
        \quad\quad
        Eric McCarthy
        \institute{Kestrel Institute \quad \url{https://kestrel.edu}}}

\def\titlerunning{Yul Formalization and Transformations}
\def\authorrunning{A. Coglio and E. McCarthy}

\maketitle

\begin{abstract}
Yul is an intermediate language used in the compilation of
the Solidity programming language for Ethereum smart contracts.
The compiler applies customizable sequences of transformations to Yul code.
To help ensure the correctness of
these transformations and their sequencing,
we used the ACL2 theorem prover to develop
a formalization of the syntax and semantics of Yul,
proofs relating static and dynamic semantics,
a formalization of some Yul code transformations,
and correctness proofs for these transformations.
\end{abstract}


\section{Introduction}
\label{sec:introduction}

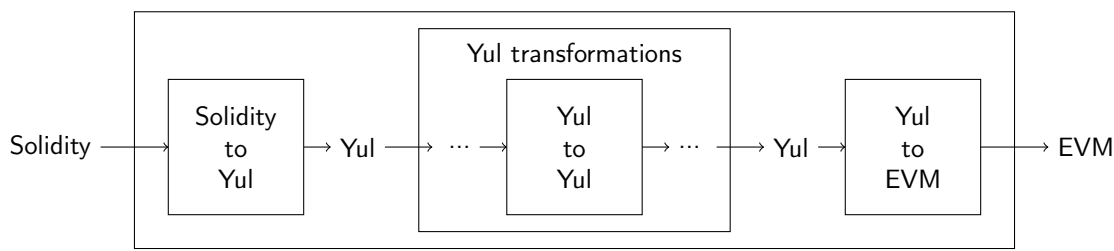
\begin{figure}
\centering

\begin{tikzpicture}[scale=0.9]
\sf
\small

\draw (-7cm, 0) node[anchor=east] {Solidity};
\draw (7cm, 0) node[anchor=west] {EVM};

\draw (-6.5cm, -1.5cm) rectangle (6.5cm, 2cm);

\draw (-1cm, -1cm) rectangle (1cm, 1cm);
\draw (0, 0) node
 {\begin{minipage}{2cm}
  \centering
  Yul \\ to \\ Yul
  \end{minipage}};

\draw (-2.3cm, -1.25cm) rectangle (2.3cm, 1.75cm);
\draw (0, 1.4cm) node {Yul transformations};

\draw (-6cm, -1cm) rectangle (-4cm, 1cm);
\draw (-5cm, 0) node
 {\begin{minipage}{2cm}
  \centering
  Solidity \\ to \\ Yul
  \end{minipage}};

\draw (4cm, -1cm) rectangle (6cm, 1cm);
\draw (5cm, 0) node
 {\begin{minipage}{2cm}
  \centering
  Yul \\ to \\ EVM
  \end{minipage}};

\draw (-3.2cm, 0) node {Yul};
\draw (3.2cm, 0) node {Yul};

\draw[->] (-7cm, 0) -- (-6cm, 0);
\draw[->] (6cm, 0) -- (7cm, 0);

\draw[->] (-4cm, 0) -- (-3.6cm, 0);
\draw[->] (3.6cm, 0) -- (4cm, 0);

\draw[->] (-2.8cm, 0) -- (-2.1cm, 0);
\draw[->] (2.1cm, 0) -- (2.8cm, 0);

\draw (-1.7cm, 0) node {...};
\draw (1.7cm, 0) node {...};

\draw[->] (-1.4cm, 0) -- (-1cm, 0);
\draw[->] (1cm, 0) -- (1.4cm, 0);

\end{tikzpicture}

\caption{Solidity Compiler with Yul Transformations}
\label{fig:compiler}
\end{figure}

Solidity \cite{solidity-www,solidity-doc} is a programming language for
writing smart contracts for the Ethereum blockchain \cite{ethereum-www}.
Solidity is compiled to
EVM (Ethereum Virtual Machine) bytecode \cite{yellow-paper},
which is directly executed by transactions on the blockchain.
The Solidity compiler includes the preferred option
to translate Solidity to EVM bytecode
via the intermediate language Yul \cite{yul-language}
(see \figref{fig:compiler}):
first, Solidity is turned into Yul
with a relatively simple translation;
next, the Yul code undergoes several optimizing transformations;
finally, the optimized Yul code is turned into EVM bytecode
with another relatively simple translation.
The rationale is to move most of the compilation complexity
into the Yul code transformations,
which eases the task because
Yul is simpler than Solidity and more structured than EVM bytecode.
Yul is also used to write inline assembly in Solidity,
i.e.\ to embed EVM bytecode (with Yul syntax) directly in the Solidity code,
which is sometimes necessary in Ethereum smart contracts.

Yul is designed to be usable as an intermediate language to compile
other front ends than Solidity to other back ends than EVM bytecode.
In line with this aspiration,
Yul consists of a core (`pure Yul' or `generic Yul')
independent from front and back ends,
which is extensible with dialects
tailored to specific front and back ends.
Currently only the EVM dialect is defined,
for compiling Solidity to EVM bytecode.
A Yul dialect extends the Yul core with specific types and operations.

A few tens of Yul transformations have been defined and implemented
\cite{yul-transformations}.
Some are dialect-independent, while others are EVM-dialect-specific.
Some transformations assume that others have already taken place,
i.e.\ they expect the code to be in a certain form,
which the previous transformations produce.
Some transformations, or sequences of transformations,
may be iterated, i.e.\ applied multiple times until
either nothing changes or an iteration limit is reached.
The Solidity compiler uses a default sequence of transformations,
which can be overridden by the user.

It is critical that Yul transformations and their sequencing are correct.
Each transformation must be applied to code of the expected form,
as produced by the preceding transformations,
and must produce code semantically equivalent to the input code.
This paper reports on our preliminary work towards addressing these problems,
using the ACL2 theorem prover \cite{acl2-www}.
We developed a formalization of the syntax and semantics of Yul,
which covers the generic core and a small portion of the EVM dialect;
we proved some properties of the formalization,
most notably that the static semantic checks
rule out dynamic semantic errors.
We formalized some Yul transformations,
and verified that they preserve both static and dynamic semantics.
We formalized some restrictions on Yul code
that are expected by some transformations,
and verified that they are preserved by some transformations.
Although we have only scratched the surface of
verifying transformations and their sequencing,
we believe that our work shows the feasibility of the approach.
Our ACL2 library for Yul, which contains our development,
is available at \citecode{books/kestrel/yul}{[books]/kestrel/yul}
and documented at \citeman{YUL____YUL}{yul}.

After providing some background on Yul in \secref{sec:background},
we describe our formalization of Yul in \secref{sec:formalization},
and our work on verified transformations in \secref{sec:transformations}.
Related work is surveyed in \secref{sec:related},
while future work is discussed in \secref{sec:future}.
Some closing remarks are given in \secref{sec:conclusion}.

Background on Solidity and the EVM can be found in a variety of sources,
starting with the Ethereum web site \cite{ethereum-www}.
However, knowledge of Solidity and the EVM is not required to read this paper.

\section{Background on Yul}
\label{sec:background}

Yul is a statically typed, block-structured, imperative language.
Statements consist of
function definitions,
variable declarations and assignments,
conditionals,
loops,
control transfers,
expressions (for side effects),
and nested blocks.
Expressions consist of literals, variables, and function calls.
A function takes zero or more inputs
and returns zero or more outputs;
if it returns no outputs, it is only used for side effects, as a statement.

The Yul core has no types, and no syntax to define types.
The EVM dialect has a single type \code{u256},
consisting of 256-bit unsigned integers.
If a type is omitted (e.g.\ in a variable declaration),
it defaults to a type specified by the dialect;
the EVM dialect necessarily defaults to \code{u256},
which is the only type.

The Yul core includes boolean literals (\code{true} and \code{false}),
numeric literals in decimal and hexadecimal base
(e.g.\ \code{64738} and \code{0xff0012}),
and certain forms of string literals
(e.g.\ \code{"abc"} and \code{hex"90a4"}).
The EVM dialect defines the meaning of literals as \code{u256} values:
boolean literals denote 0 and 1,
numeric literals denote the obvious,
and string literals yield byte sequences
interpreted as integers in base 256.
It is a static error if a literal denotes $2^{256}$ or more,
e.g.\ if a string literal yields more than 32 bytes.

Variables are declared via the \code{let} keyword,
with or without a type (see above),
and with or without an initializing expression;
if the latter is missing, the variable is initialized to \code{0}.
In the EVM dialect,
\code{let~x} declares
a variable \code{x} of type \code{u256} initialized to \code{0},
while \code{let~y~:=~x} declares a variable \code{y} of type \code{u256}
initialized to the current value of \code{x}.
A variable assignment is like a variable declaration without \code{let},
e.g.\ \code{x~:=~17} assigns \code{17} to \code{x}.
A call of a function that returns two or more outputs can be used
to initialize, or to assign to, multiple variables,
e.g.\ \code{let~a,~b~:=~f(...)} initializes
\code{a} and \code{b} to the first and second result of \code{f}.
If multiple variables are declared without initializing expressions,
they are all initialized to \code{0}, as in the case of a single variable,
e.g.\ \code{let~a,~b}.

A function definition returns results by assigning them to its output variables,
which are declared as part of the function definition,
along with the input variables.
For example, in a function definition of the form
\code{function~f(x,~y)~->~a,~b~\{~...~a~:=~...~b~:=~...~\}},
the input variables are \code{x} and \code{y},
and the output variables are \code{a} and \code{b}.
If a function terminates execution
without assigning a value to some output variables,
the corresponding result is \code{0}.
The \code{leave} statement can be used to return from a function,
whose execution otherwise terminates at the end of its body block.

The Yul core has no built-in functions.
The EVM dialect provides several tens of built-in functions,
corresponding to EVM bytecode instructions.
For example, the function \code{add},
which takes two \code{u256} inputs
and returns their \code{u256} sum (modulo $2^{256}$),
corresponds to the EVM bytecode instruction \code{ADD}.
An assignment like \code{z~:=~add(x, y)} in inline assembly
represents, and is translated to (as the last compilation step),
an \code{ADD} instruction in EVM bytecode.
Although the EVM is stack-based,
Yul is essentially register-based (where the registers are the variables);
the rationale is to facilitate understanding and manipulation of Yul code,
while keeping the translation to EVM bytecode still relatively simple.

\code{for} loops are structurally similar to C and Java:
there is an initialization block,
a test expression,
an update block,
and a body block;
the \code{break} and \code{continue} statements
can be used to break out of a loop or to skip the rest of an iteration.
\code{if} conditionals have a `then' branch and no `else' branch.
\code{switch} conditionals have
\code{case} branches based on literals,
and optional \code{default} branches.
There are no go-to statements;
the rationale for this, and for having structured control flow,
is to facilitate understanding and manipulation of Yul code.

Blocks are delimited by curly braces, i.e.\ \code{\{~...~\}},
as in many other languages.
Statements are not terminated by semicolons;
note that there are no infix operators, only function calls,
which makes parsing easier.
There are line comments \code{//~...}
and block comments \code{/*~...~*/},
as in many other languages.

The Yul documentation \cite{yul-language}
includes a grammar, as is customary,
and a semi-formal semantics, which is much less customary.
The latter is an evaluation function over the Yul syntactic constructs,
written in a mix of mathematics, pseudo-code, and English prose.

\section{Yul Formalization}
\label{sec:formalization}

Our formalization covers the Yul core and a small portion of the EVM dialect.
Extending it to cover the whole EVM dialect,
and generalizing it to accommodate other dialects,
are both future work.

\subsection{Abstract Syntax}
\label{sec:abstract-syntax}

The abstract syntax is the fulcrum of our development:
concrete syntax abstracts to it;
static and dynamic semantics are defined on it;
and transformations manipulate it.
For defining static and dynamic semantics,
the abstract syntax could abstract away
all the concrete syntax information that does not affect said semantics.
But for defining transformations,
it is beneficial to retain enough concrete syntax information
to reduce incidental differences between code before and after transformations,
to facilitate inspection and debugging;
however, retaining excessive concrete syntax information
may add complexity without significant additional benefit.
In formalizing the abstract syntax,
we tried to strike the right balance:
we keep all the syntactic details of literals,
but we drop whitespace and comments.

The abstract syntax is formalized as a collection of algebraic data types,
using the fixtype library \cite{fty-paper} \citeman{ACL2____FTY}{fty}.
For example, expressions are formalized as
\begin{bcode}
(fty::deftagsum expression
  (:path ((get path)))
  (:literal ((get literal)))
  (:funcall ((get funcall)))
  :pred expressionp)
\end{bcode}
i.e.\ an expression is either a path, or a literal, or a function call;
a path is a sequence of identifiers separated by dots,
which are used as variable names.%
\footnote{The motivation for using paths as variable names seems to be that
Yul variables may represent nested fields in Solidity.}
As another example, statements are formalized as
\begin{bcode}
(fty::deftagsum statement
  (:block ((get block)))
  (:variable-single ((name identifier) (init expression-option)))
  (:variable-multi ((names identifier-list) (init funcall-optionp)))
  (:assign-single ((target path) (value expression)))
  (:assign-multi ((targets path-list) (value funcall)))
  (:funcall ((get funcall)))
  (:if ((test expression) (body block)))
  (:switch ((target expression) (cases swcase-list) (default block-option)))
  (:for ((init block) (test expression) (update block) (body block)))
  (:break ())
  (:continue ())
  (:leave ())
  (:fundef ((get fundef)))
  :pred statementp)
\end{bcode}
i.e.\ a statement is either a block,
or a (single or multiple) variable declaration,
or a (single or multiple) variable assignment,
or a function call,
or an (\code{if} or \code{switch}) conditional,
or a (\code{for}) loop,
or a (\code{break} or \code{continue} or \code{leave}) control transfer,
or a function definition.
Overall, the definition of the abstract syntax is unremarkable,
directly derived from the concrete syntax.

\subsection{Concrete Syntax}
\label{sec:concrete-syntax}

To formalize the concrete syntax,
we developed an ABNF \cite{abnf-rfc,abnf-string-rfc} grammar of Yul,
as a straightforward transcription of the grammar in
\cite{yul-language,solidity-doc}.%
\footnote{As explained in
\citeman{YUL____CONCRETE-SYNTAX}{yul::concrete-syntax},
\cite{yul-language,solidity-doc} contains an old grammar and a new grammar,
both of which we have transcribed to ABNF.
This paper focuses on the new grammar,
for which we have also developed a parser.}
For example, the ABNF grammar rule for expressions is
\begin{bcode}
expression = path / literal / function-call
\end{bcode}
and the ABNF grammar rule for statements is
\begin{bcode}
statement = block / variable-declaration / assignment / function-call
          / if-statement / switch-statement / for-statement
          / %s"leave" / %s"break" / %s"continue" / function-definition
\end{bcode}
which also show the correspondence with
the examples in \secref{sec:abstract-syntax}.%
\footnote{While the abstract syntax of statements has
different cases for single and multiple variable declarations and assignments,
the grammar makes that distinction in the rules for
\code{variable-declaration} and \code{assignment} (not shown here).}
The verified ABNF grammar parser
\cite{abnf-paper} \citeman{ABNF____GRAMMAR-PARSER}{abnf::grammar-parser}
turns the ABNF grammar of Yul into an ACL2 representation with formal semantics,
according to the formalization of the ABNF notation
\cite{abnf-paper} \citeman{ABNF____NOTATION}{abnf::notation}.

As is customary in programming languages, the grammar consists of
a lexical sub-grammar,
which specifies how sequences of characters are organized into lexemes
(i.e.\ tokens, whitespace, and comments),
and a syntactic sub-grammar,
which specifies how tokens (after discarding whitespace and comments)
are organized into expressions, statements, and related constructs.
As is also customary, the lexical sub-grammar is further constrained
by taking the longest possible lexeme at each point
(e.g.\ \code{xy} is a single lexeme, not two lexemes \code{x} and \code{y}),
and by the fact that keywords are not identifiers.
A complete formalization of the concrete syntax
should include these restrictions,
but this is future work.

We developed an (unverified) executable parser of Yul in ACL2.
The lexer is partially generated,
via some preliminary ABNF parser generation tools
\citeman{ABNF____DEFDEFPARSE}{abnf::defdefparse};
the rest is handwritten, closely following the lexical grammar.
The parser proper is handwritten, closely following the syntactic grammar,
according to a recursive descent strategy.

\subsection{Static Semantics}
\label{sec:static-semantics}

The static semantics consists of
efficiently checkable restrictions on the syntax,
informally stated in \cite{yul-language}.
An example is that a function must be called with the right number of arguments.
The Solidity compiler must:
enforce these restrictions on inline assembly;
translate Solidity code to Yul code that satisfies these restrictions;
and transform Yul code preserving these restrictions.
The static semantics is formalized as
executable ACL2 functions that check these restrictions
recursively on the abstract syntax.

The Yul scoping rules involve the notions of visibility and accessibility,
which differ between functions and variables.
They are explained below, using the following code as example:
\begin{bcode}
\{ // block 1:
  let x
  function f () \{ // block 2:
    function h () \{ ... \}
    let y
    \{ // block 3:
      let z
    \}
  \}
  function g () \{ // block 4:
    let y
    function h () \{ ... \}
  \}
\}
\end{bcode}

A function is both visible and accessible
in the whole block where its definition occurs,
even before the definition,
and including all the nested blocks.
In the example above:
\code{f} and \code{g} are visible and accessible
everywhere in block 1 (including blocks 2, 3, and 4),
but not outside block 1;
the \code{h} defined in \code{f} is visible and accessible
everywhere in block 2 (including block 3),
but not outside block 2 (e.g.\ not in block 4);
and so on.
A function definition is disallowed if
the function name is already visible and accessible,
e.g.\ no function \code{g} can be declared in \code{f}.
The \code{h} defined in \code{f}
is distinct from
the \code{h} defined in \code{g}.

A variable is visible
from just after its declaration
to the end of the block where it occurs,
including all the nested blocks;
the variable is accessible in the same portion of the block,
except in nested functions.
In the example above:
\code{x} is visible
in the portion of block 1 just after its declaration,
including blocks 2, 3, and 4,
but it is not accessible in blocks 2, 3, and 4;
the \code{y} declared in \code{f} is visible
in the portion of block 2 just after its declaration,
including block 3,
and it is also accessible in block 3;
and so on.
A variable declaration is disallowed if
the variable name is already visible,
regardless of whether it is also accessible.
The \code{y} declared in \code{f}
is distinct from
the \code{y} declared in \code{g}.

Visibility means lexical scoping,
i.e.\ which names can be seen from where,
while accessibility means the ability to reference those seen names.
When a function is called, a fresh variable area is created,
without the ability to reference the variables of the caller:
this is why accessibility of variables stops at function boundaries,
and why the notions of visibility and accessibility differ for variables.
For functions, visibility and accessibility coincide.

Our formalized static semantics checks the above scoping rules,
using symbol tables for variables and functions.
Since neither the Yul core nor the EVM dialect have syntax for types,
symbol tables for variables are just finite sets of variable names
(all of which have the same type),
while symbol tables for functions are finite maps
from function names to function ``types'',
where the latter are pairs $(n,m)$
where $n$ is the number of inputs and $m$ is the number of outputs.
A function must be called with $n$ arguments;
the call must be a statement if $m = 0$ (for side effects),
or used to initialize or assign $m$ variables if $m \neq 0$.

Restrictions on where \code{break}, \code{continue}, and \code{leave} may occur
are enforced by calculating and checking
the possible ways in which statements and blocks may terminate.
There are four possible ways, called `modes'
(also used in the dynamic semantics; see \secref{sec:dynamic-semantics}):
three modes corresponding to those three statements,
and one mode corresponding to `regular' termination.

Most static semantic checks are dialect-independent,
except that literals are interpreted as denoting \code{u256} values,
which are thus checked to be below $2^{256}$;
this is EVM-dialect-specific.
Our static semantics provides the option
to initialize the function symbol table
with the types of the built-in functions of the EVM dialect,
so that Yul code in the EVM dialect can be properly checked.

\subsection{Dynamic Semantics}
\label{sec:dynamic-semantics}

The dynamic semantics is formalized as
a defensive big-step executable interpreter of the abstract syntax.
Each call of an ACL2 function of the interpreter attempts to execute
its input abstract syntax construct completely,
recursively executing the sub-constructs.
Since the execution of certain constructs may not terminate,
the ACL2 functions take, as additional input, an artificial counter
that limits the depth of the mutual recursion:
the counter is decremented by one at each recursive call,
and used as measure, making the termination proof straightforward.
The interpreter is defensive in the sense that
it checks the necessary safety conditions,
e.g.\ that each function is called with the right number of arguments,
without relying on the static semantics
(see \secref{sec:static-soundness}
for the relation between static and dynamic semantics).

This approach matches the semi-formal semantics in \cite{yul-language},
which is also a big-step interpreter.
Besides the syntactic construct (expression, statement, etc.),
the interpreter takes as input, and returns as output,
a global state $G$, which is dialect-specific,
and a local state $L$, which is dialect-independent;
the interpreter also returns one of the four termination modes
described in \secref{sec:static-semantics}.
$L$ is the state of the local variables.
In the EVM dialect, $G$ consists of various areas of memory,
and provides read access to some blockchain state (e.g.\ current block number).
Our ACL2 interpreter has the same structure.

The ACL2 function to execute statements is
\begin{bcode}
(define exec-statement
  ((stmt statementp) (cstate cstatep) (funenv funenvp) (limit natp))
  :returns (outcome soutcome-resultp)
  (b* (((when (zp limit)) ...)) ; return limit error
    (statement-case stmt
                    :block (exec-block stmt.get cstate funenv (1- limit))
                    :leave (make-soutcome :cstate cstate :mode (mode-leave))
                    ...)) ; handle the other kinds of statement
  :measure (nfix limit))
\end{bcode}
where:
\begin{itemize}[nosep]
\item
\code{stmt} is the statement,
which is handled by cases
(see the type definition in \secref{sec:abstract-syntax}).
\item
\code{cstate} is a computation state,
which wraps a finite map from variable names to 256-bit unsigned integers,
modeling the local state $L$,
tailored to the EVM dialect because values have type \code{u256}:
\begin{bcode}
(fty::defprod cstate
  ((local lstate)) ; finite map from identifiers to values
  :pred cstatep)
\end{bcode}
We do not yet model the global state $G$ (which is complex),
but the reason to wrap the type \code{lstate} into \code{cstate}
is to accommodate the future addition of a \code{(global gstate)} component.
The local state is a flat map, not a stack of maps corresponding to scopes,
because each called function starts a new local state
(consisting of the function's input and output variables),
and because nested block scopes cannot shadow variables.
\item
\code{funenv} is a function environment,
i.e.\ a stack of finite maps from function names to
function information consisting of inputs, outputs, and body:
\begin{bcode}
(fty::defprod funinfo
  ((inputs identifier-list)
   (outputs identifier-list)
   (body block))
  :pred funinfop)
\end{bcode}
Unlike the local state,
the function environment is a stack because of
the different scoping rules of functions compared to variables:
the stack corresponds to the lexical scoping of functions;
when a function is called,
the stack is trimmed down to the scope of that function.
With reference to the example code in \secref{sec:static-semantics}:
when executing the \code{h} defined in \code{g},
the function environment contains
a scope for block 1 with \code{f} and \code{g},
a scope for block 4 with \code{h},
and a scope for the body of \code{h};
if \code{h} calls \code{f},
the two top scopes are popped, leaving only the one for block 1,
because \code{f} can only access the functions in that scope.
\item
\code{limit} is the artificial counter,
which \code{exec-statement} tests with \code{zp} as first thing,
returning an error value indicating that the limit is exhausted
if that is the case.
This \code{limit} is decremented at each recursive call,
e.g.\ in the call of \code{exec-block},
and used as measure for the mutual recursion.
\item
\code{outcome} is either an error value or a statement outcome of type
\begin{bcode}
(fty::defprod soutcome
  ((cstate cstate)
   (mode mode))
  :pred soutcomep)
\end{bcode}
which consists of a possibly updated computation state
and a mode of termination.
Inspired by the \code{Result} type in Rust,
the type \code{soutcome-result} extends \code{soutcome} with error values.
\item
A block statement is executed by executing the block
with a separate function \code{exec-block},
which extends the function environment with a new scope,
executes the statements in the block,
and then pops the function environment
and reduces the local state to the variables before the block
(the function \code{exec-block} is not shown here).
\item
The execution of a \code{leave} statement
returns the \code{leave} termination mode
without changing the computation state.
That termination mode is propagated upwards,
and treated the same as regular termination
by the ACL2 function \code{exec-function} that executes Yul functions.
\item
The code to execute the other kinds of statements is not shown.
\end{itemize}

The execution of expressions returns an error value
or an expression outcome of type
\begin{bcode}
(fty::defprod eoutcome
  ((cstate cstate)
   (values value-list))
  :pred eoutcomep)
\end{bcode}
which is analogous to statement outcomes,
but with a list of zero or more values instead of a termination mode.
Although function calls, and thus expressions,
have no side effects on the local state,
because each function has its own local state,
our ACL2 interpreter accommodates
the extension with side effects on the global state,
since expression outcomes include a possibly updated computation state.

Besides returning an error value when the artificial limit is exhausted,
our ACL2 interpreter returns an error value
when a defensive check fails,
e.g.\ a referenced variable or function is not accessible,
a \code{break} is executed outside a loop body,
a function is given the wrong number of arguments,
etc.

Except for using values of type \code{u256},
the current interpreter has no support for the EVM dialect.
Adding support involves modeling the global state $G$
and modeling the EVM built-in functions
via ACL2 code that manipulates the global state.
Adding this support is future work.

\subsection{Static Soundness}
\label{sec:static-soundness}

We proved the soundness of the static semantics
with respect to the dynamic semantics:
if the checks of the static semantics are satisfied,
the dynamic semantics never returns an error value,
except when the artificial limit is exhausted.
In other words, the defensive checks of the dynamic semantics
are guaranteed to succeed if the checks of the static semantics succeed.
This kind of property provides a major validation of
the design and formalization of a programming language.
The converse property, i.e.\ static completeness,
namely that if the dynamic semantics never returns error values
(except for exhausting the artificial limit)
then the static semantics succeeds,
cannot hold for any decidable static semantics,
because it is undecidable whether the dynamic semantics returns error values.%
\footnote{For instance, in a conditional statement \code{if~E~B},
where \code{E} is an expression and \code{B} is a block,
\code{B} may have a static error
that never causes a dynamic error due to \code{E} being always false.
In general, it may be possible to prove a static completeness property
with respect to an extended dynamic semantics
that nondeterministically chooses among all possible branches
regardless of the actual values of the tests that control branching,
as done in \cite{java-subroutines}.
In \code{if~E~B}, after evaluating \code{E},
that extended dynamic semantics would nondeterministically
either execute \code{B} or skip it, regardless of the value of \code{E}.
As in \cite{java-subroutines},
this hypothetical static completeness property for Yul
would show that the checks of the static semantics are, in a sense,
the most liberal possible.}

This formulation of static soundness relies on the fact that,
in the Yul core covered by our dynamic semantics
(see \secref{sec:dynamic-semantics}),
the only error values are for the exhaustion of the artificial limit
and for defensive checks also checked by the static semantics.
If some dialect-specific built-in function
can fail in ways not detectable by the static semantics
(e.g.\ division by zero),
then static soundness should be reformulated
to rule out only the errors detectable by the static semantics.

The static semantics involves function and variable symbol tables,
while the dynamic semantics involves
function environments and computation states.
To formulate static soundness, those are related as follows:
a function environment abstracts to a function symbol table,
merging the scopes
and only keeping the numbers of inputs and outputs of each function;
and a computation state abstracts to a variable symbol table,
keeping only the variables in the domain of the local state map.

There is a static soundness theorem for each mutually recursive ACL2 function,
proved by induction on the mutual recursion.
The theorem for expression execution is
\begin{bcode}
(defthm exec-expression-static-soundness
  (b* ((results (check-safe-expression
                 expr (cstate-to-vars cstate) (funenv-to-funtable funenv)))
       (outcome (exec-expression expr cstate funenv limit)))
    (implies (and (funenv-safep funenv)
                  (not (reserrp results))
                  (not (reserr-limitp outcome)))
             (and (not (reserrp outcome))
                  (equal (cstate-to-vars (eoutcome->cstate outcome))
                         (cstate-to-vars cstate))
                  (equal (len (eoutcome->values outcome))
                         results)))))
\end{bcode}
where:
\begin{itemize}[nosep]
\item
\code{cstate-to-vars} abstracts a computation state
to a variable symbol table.
\item
\code{funenv-to-funtable} abstracts a function environment
to a function symbol table.
\item
\code{check-safe-expression}, from the static semantics,
checks the safety of the expression \code{expr},
returning the number of results if successful,
or an error value otherwise.
\item
\code{exec-expression}, from the dynamic semantics,
executes the expression \code{expr},
returning an expression outcome, described in \secref{sec:dynamic-semantics}.
\item
The theorem assumes that:
the functions in the function environment
pass the checks of the static semantics,
formalized by \code{funenv-safep};
\code{results} is not an error value (recognized by \code{reserrp}),
i.e.\ \code{check-safe-expression} succeeds,
and \code{results} is the number of results of \code{expr};
the execution of \code{expr} does not exhaust the artificial limit,
where \code{reserr-limitp} recognizes
error values that come from exhausting the limit.
\item
The theorem concludes that:
the execution of \code{expr} does not return an error value;
the new computation state abstracts to the same variable symbol table
as the old computation state;
the actual number of values in the outcome
coincides with the statically computed one.
\end{itemize}
The theorems for the other \code{exec-...} functions are similar;
when statement outcomes are involved instead of expression outcomes,
the conclusion about the number of values instead says that
the actual termination mode is an element of
the set of possible modes calculated by the static semantics.

In these theorems, the \code{funenv-safep} hypothesis is critical
to establish the static safety hypotheses for
the code of each called Yul function,
which is obtained from the function environment.
As the function environment is extended,
the preservation of its safety relies on the static safety hypotheses
for the code containing the Yul function definitions added to the environment.
When the function environment is trimmed,
its safety is preserved because it applies element-wise.

The inductively proved theorems about the \code{exec-...} functions
are preceded by, and rely on, the proofs of several theorems about
\code{funenv-safep}, \code{funenv-to-funtable}, and other ACL2 functions.
Overall, the proofs are not conceptually difficult, but involve a bit of work.

\section{Yul Transformations}
\label{sec:transformations}

While the formalization of Yul described in \secref{sec:formalization}
has value on its own,
our primary motivation for developing it was to support
the verification of Yul code transformations in the Solidity compiler.
Our work on transformations is fairly preliminary, yet illustrative.

\subsection{Approach}
\label{sec:approach}

Since the Solidity compiler is written in C++ \cite{solidity-compiler},
verifying the implementation of Yul transformations is a daunting proposition.
It is more feasible to generate,
each time a Yul transformation is run,
a proof of the correctness of the new code with respect to the old code;
a verifying (not verified) compiler approach.

Extending the Solidity compiler to generate such proofs is impractical,
due to its complexity and ownership.
A more viable approach is to
(1) replicate the Yul transformations in ACL2,
(2) verify correctness properties of the replicated transformations, and
(3) validate the replicated transformations
by checking that they are consistent with the transformations in the compiler.
Performing the third step every time a transformation is run,
and instantiating the general theorems of the second step to the run,
achieves the same goal as a proof-generating extension of the compiler,
but without modifying the compiler;
in \cite{zkcirc-paper},
we introduced the term `detached proof-generating extension'
for this approach.

\begin{figure}
\centering

\begin{tikzpicture}[scale=0.9]
\sf
\small

\draw (0, 0) node
 {\begin{minipage}{6cm}
  \centering
  Yul transformation in ACL2
  \[
  \begin{array}{c}
   \begin{array}{ccc}
   T(x,y) \triangleq \ldots & &
   C(x,y) \triangleq \ldots
   \end{array} \\
  \vdash \forall x, y. \ T(x,y) \Longrightarrow C(x,y)
  \end{array}
  \]
  \end{minipage}};
\draw (-3.1cm, -1.2cm) rectangle (3.1cm, 1.2cm);

\draw (0, 3cm) node {Yul transformation in C++};
\draw (-3.1cm, 2.5cm) rectangle (3.1cm, 3.5cm);

\draw (0, -3cm) node { theorem generation};
\draw (-3.1cm, -3.5cm) rectangle (3.1cm, -2.5cm);

\draw (-6cm, 3cm) node {Yul};
\draw (6cm, 3cm) node {Yul};
\draw (-6cm, 0) node {JSON};
\draw (6cm, 0) node {JSON};
\draw (-6cm, -3cm) node {AST $a$};
\draw (6cm, -3cm) node {AST $b$};

\draw (-6cm, 1.5cm) node {export};
\draw (-6.8cm, 1.1cm) rectangle (-5.2cm, 1.9cm);

\draw (6cm, 1.5cm) node {export};
\draw (5.2cm, 1.1cm) rectangle (6.8cm, 1.9cm);

\draw (-6cm, -1.5cm) node {convert};
\draw (-6.8cm, -1.9cm) rectangle (-5.2cm, -1.1cm);

\draw (6cm, -1.5cm) node {convert};
\draw (5.2cm, -1.9cm) rectangle (6.8cm, -1.1cm);

\draw (0, -5cm) node {$\vdash C(a,b)$};

\draw[->] (-5.5cm, 3cm) -- (-3.1cm, 3cm);
\draw[->] (3.1cm, 3cm) -- (5.5cm, 3cm);

\draw[->] (-6cm, 2.6cm) -- (-6cm, 1.9cm);
\draw[->] (6cm, 2.6cm) -- (6cm, 1.9cm);

\draw[->] (-6cm, 1.1cm) -- (-6cm, 0.4cm);
\draw[->] (6cm, 1.1cm) -- (6cm, 0.4cm);

\draw[->] (-6cm, -0.4cm) -- (-6cm, -1.1cm);
\draw[->] (6cm, -0.4cm) -- (6cm, -1.1cm);

\draw[->] (-6cm, -1.9cm) -- (-6cm, -2.6cm);
\draw[->] (6cm, -1.9cm) -- (6cm, -2.6cm);

\draw[->] (-5.2cm, -3cm) -- (-3.1cm, -3cm);
\draw[->] (5.2cm, -3cm) -- (3.1cm, -3cm);

\draw[->] (0, -3.5cm) -- (0, -4.5cm);

\draw[->, dashed] (0cm, -1.2cm) -- (0, -2.5cm);

\end{tikzpicture}

\caption{Proof Generation Approach for Yul Transformations}
\label{fig:approach}
\end{figure}
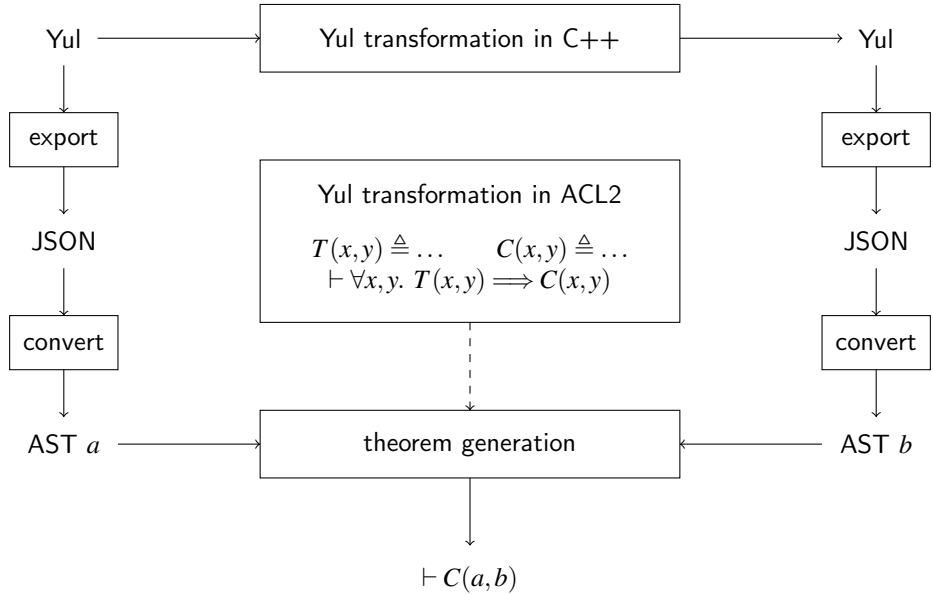

\figref{fig:approach} visualizes the approach.
The top box is any one of the Yul transformations in the Solidity compiler.
The middle box is the replicated transformation in ACL2,
formalized as a predicate $T(x,y)$ on old code $x$ and new code $y$,
accompanied by a correctness predicate $C(x,y)$
(which may vary slightly across transformations),
and a general theorem that $T$ implies $C$;
this theorem is proved once, for each transformation,
under user guidance,
with effort dependent on the complexity of the transformation.
The Solidity compiler has facilities
to output Yul abstract syntax trees in JSON format
at various stages of the compilation process:
these facilities are depicted as the `export' boxes.
We have built a tool, in ACL2, to convert the resulting JSON
into abstract syntax trees (ASTs) of our ACL2 formalization of Yul;
this is depicted as the `convert' boxes.
Each time the transformation is run,
ASTs $a$ and $b$ for the old and new code can be automatically generated;
the bottom box can check that $T(a,b)$ holds,
and if so it can instantiate the general theorem
to obtain a proof of $C(a,b)$ automatically.
These last sentences say `can' because
we have not implemented this workflow yet,
although we see no obstacle to doing that.

An alternative to verifying the correctness of the replicated transformations
is to have them generate proofs of correctness,
as done in APT \cite{apt-www,apt-simplify-paper,apt-isodata-paper}
and ATC \cite{atc-paper},
pushing the verifying compiler approach further.
But having pursued that approach in the tools just mentioned,
for Yul we wanted to explore the verification of the transformations.
In summary, our approach for Yul transformation is
a combination of verifying and verified compiler:
the former for the transformations in the Solidity compiler,
and the latter for the replicated transformations in ACL2.

\subsection{Definitions}
\label{sec:definitions}

We have formalized the
\code{ForLoopInitRewriter},
\code{DeadCodeEliminator}, and
\code{Disambiguator}
transformations \cite{yul-transformations}.
The first two are quite simple;
the third one is relatively simple,
but illustrates a general important point.

The \code{ForLoopInitRewriter} transformation
moves the initialization component of a \code{for} loop
just before the loop and wraps it and the loop in a block,
e.g.\ the loop
\begin{bcode}
for \{ <init> \} <test> \{ <update> \} \{ <body> \}
\end{bcode}
is transformed into the block%
\footnote{More precisely, the transformation is recursively applied to
\code{<init>}, \code{<test>}, \code{<update>}, and \code{<body>} as well.}
\begin{bcode}
\{ <init> for \{ \} <test> \{ <update> \} \{ <body> \} \}
\end{bcode}
The scoping rules for \code{for} loops involve an exception:
the scope of variables declared and functions defined
in the initialization block
extends to the whole loop (test, update, and body).
The purpose of performing this transformation
is to obviate the need for successive transformations
to deal with this scoping exception.
This transformation is easily defined in ACL2,
by recursion on the abstract syntax.

The \code{DeadCodeEliminator} transformation
removes a simple form of dead code,
namely the code in a block that follows
a \code{break}, \code{continue}, or \code{leave} statement,
e.g.\ the block
\begin{bcode}
\{ <live> break <dead> \}
\end{bcode}
is transformed into the block%
\footnote{More precisely, the transformation is recursively applied to
\code{<live>} as well.}
\begin{bcode}
\{ <live> break \}
\end{bcode}
The purpose of this transformation is to reduce code size.

The \code{Disambiguator} transformation
makes all the variable and function names unique across the whole program.
For instance, the example code in \secref{sec:static-semantics}
is transformed into something like
\begin{bcode}
\{
  let x
  function f () \{
    function h1 () \{ ... \}
    let y1
    \{
      let z
    \}
  \}
  function g () \{
    let y2
    function h2 () \{ ... \}
  \}
\}
\end{bcode}
where the two different variables \code{y} and functions \code{h}
are renamed apart.
There are many ways to rename variables and functions apart,
differing in the exact choice of names.
To make our ACL2 definition of the transformation simpler,
and independent from the choice of names made by the Solidity compiler,
we formalized it as a relation instead of a function:
the relation holds on old and new code exactly when
they are the same except for a consistent renaming of variables and functions
such that the new code has globally unique names.
In fact, our definition consists of four independent components:
\begin{itemize}[nosep]
\item
A binary relation expressing consistent variable renaming.
\item
A binary relation expressing consistent function renaming.
\item
A unary relation expressing global uniqueness of variable names.
\item
A unary relation expressing global uniqueness of function names.
\end{itemize}

The binary relation for variable renaming
consists of a family of recursive functions, such as
\begin{bcode}
(define statement-renamevar ((old statementp) (new statementp) (ren renamingp))
  :returns (new-ren renaming-resultp)
  (statement-case
   old
   :block
   (b* (((unless (statement-case new :block)) ...) ; return error
        ((statement-block new) new)
        ((okf &) (block-renamevar old.get new.get ren)))
     (renaming-fix ren))
   :variable-single
   (b* (((unless (statement-case new :variable-single)) ...) ; return error
        ((statement-variable-single new) new)
        ((okf &) (expression-option-renamevar old.init new.init ren)))
     (add-var-to-var-renaming old.name new.name ren))
   ...)) ; handle the other kinds of statements
\end{bcode}
where:
\begin{itemize}[nosep]
\item
\code{old} and \code{new} are
the statements before and after the transformation.
\item
\code{ren} is a renaming,
i.e.\ a list of \code{cons} pairs of identifiers
\code{((x1~.~y1)~(x2~.~y2)~...)}
where \code{x1}, \code{x2}, etc.\ are all distinct
and where \code{y1}, \code{y2}, etc.\ are all distinct.
It is an injective alist with unique keys,
invertible into \code{((y1~.~x1)~(y2~.~x2)~...)}.
\code{x1}, \code{x2}, etc.\ are the variables in scope for \code{old};
\code{y1}, \code{y2}, etc.\ are the variables in scope for \code{new}.
\code{x1} in \code{old} is renamed to \code{y1} in \code{new},
\code{x2} in \code{old} is renamed to \code{y2} in \code{new},
etc.;
but \code{x1} and \code{y1} could be the same,
or \code{x2} and \code{y2} could be the same,
etc.
\item
If \code{old} is a block statement,
\code{new} must be a block statement too;
otherwise \code{statement-renamevar} returns an error value,
because \code{new} is not a valid result of transforming \code{old}.
\item
The block in \code{new} must be
a valid result of transforming the block in \code{old},
which is checked by the mutually recursive companion function
\code{block-renamevar}.
\item
Since a block contributes no new variables outside it,
\code{statement-renamevar} returns \code{ren}.%
\footnote{The fixer \code{renaming-fix} is a no-op
under the guard \code{(renamingp~ren)},
but it makes the return theorem unconditional.}
\item
If \code{old} is a (single) variable declaration, say for \code{x},
\code{new} must be one too, say for \code{y}.
The optional initializing expression of \code{new}
must be a valid result of transforming the one of \code{old},
via the renaming \code{ren},
which is then extended with the pair \code{(x~.~y)},
and returned.
\item
The code for the other kinds of statements is not shown.
\end{itemize}
The binary relation for function renaming is defined similarly.

The unary relations for unique variable and function names
go through the abstract syntax,
keep track of the set of all the names encountered so far,
whether visible/accessible or not,
and check that every variable declaration or function definition
introduces a name not already in the set.

The purpose of \code{Disambiguator}
is to make it easier for subsequent transformations
to move code around, without worrying about name conflicts.

Our formalization of \code{Disambiguator},
unlike our formalizations of the other transformations,
does not consist of executable ACL2 code to run the transformation;
it consists of executable ACL2 code to check whether
the result of the transformation is valid.
According to the approach described in \secref{sec:approach},
the purpose of formalizing the transformation in ACL2
is to verify that, every time the Solidity compiler runs it,
the new code is equivalent to the old code.
The old and new code are given as inputs to
our formalization of the transformation
to check that the action of the Solidity compiler matches our formalization;
given a general proof of the correctness of our formalization
(see \secref{sec:proofs}),
it is possible to obtain a proof of
the correctness of that run of the transformation by the Solidity compiler.
This relational approach,
which isolates the formal definition and proofs
from changeable details of the implementation,
may also be applicable to other Yul code transformations.

\subsection{Restrictions}
\label{sec:restrictions}

Some transformations expect the code to satisfy certain restrictions,
which must be established by preceding transformations,
and must be generally preserved by subsequent transformations.
We formalized some of these restrictions in ACL2
as predicates on the abstract syntax.

We formalized the restriction that
\code{for} loops have empty initialization blocks.
As mentioned in \secref{sec:definitions},
this restriction is established by \code{ForLoopInitRewriter}.

We formalized the restriction that code has no function definitions.
Some transformations (not described in this paper)
move all the function definitions (after disambiguation)
to a new top-level block,
so that subsequent transformations
can take all the function definitions from the top-level block
without worrying about nested function definitions.
Our formalized restriction just says that there are no function definitions:
it applies to the non-top-level code,
which is recursively processed by transformations,
stripped of the function definitions at the top level.

\subsection{Proofs}
\label{sec:proofs}

We proved that (our formalization of) \code{DeadCodeEliminator}
preserves the two restrictions in \secref{sec:restrictions}.
Removing code does not introduce function definitions
or code in \code{for} loop initialization blocks.
These proofs are automatic, after enabling the involved functions.

We proved that \code{DeadCodeEliminator} preserves the static semantics:
if the old code is safe, so is the new code,
assuming the restriction about the absence of function definitions.
The latter is critical:
if a function definition follows a \code{break},
removing the code after the \code{break} removes that function definition,
which the old code may be calling in non-dead code before the \code{break}.
The theorem for statements is
\begin{bcode}
(defthm check-safe-statement-of-statement-dead
  (implies (and (statement-nofunp stmt)
                (statement-noloopinitp stmt))
           (b* ((varsmodes (check-safe-statement stmt varset funtab))
                (varsmodes-dead (check-safe-statement (statement-dead stmt)
                                                      varset
                                                      funtab)))
             (implies (not (reserrp varsmodes))
                      (and (not (reserrp varsmodes-dead))
                           (equal (vars+modes->vars varsmodes-dead)
                                  (vars+modes->vars varsmodes))
                           (set::subset (vars+modes->modes varsmodes-dead)
                                        (vars+modes->modes varsmodes)))))))
\end{bcode}
where:
\begin{itemize}[nosep]
\item
\code{statement-dead} removes dead code from a statement,
according to the transformation:
\code{stmt} is the old statement,
and \code{(statement-dead~stmt)} is the new statement.
\item
\code{varset} and \code{funtab} are the variable and function symbol tables.
For this transformation, they are the same for old and new code;
more complex transformations may require transforming these tables as well.
\item
\code{check-safe-statement}, from the static semantics,
checks the safety of a statement;
if successful, it returns an updated variable symbol table,
and a set of possible termination modes.
\item
The theorem assumes that:
the old statement has no function definitions (critical hypothesis);
the old statement has empty \code{for} loop initialization blocks
(which slightly simplifies the proof);
the old statement is safe, i.e.\ \code{varmodes} is not an error value.
\item
The theorem concludes that:
the new statement is safe, i.e.\ \code{varmodes-dead} is not an error value;
the updated variable table after the new statement
is the same as the one after the old statement;
the termination modes of the new statement are a subset of
the ones of the old statement
(because the static semantics over-approximates them;
for example, if a \code{leave} followed a \code{break} in the old code,
it would be absent in the new code).
\end{itemize}
The proof is automatic, after enabling the involved functions
and also adding an \code{:expand} hint.%
\footnote{Without the \code{:expand} hint, the proof fails,
presumably due to heuristics about opening recursive functions.}

We proved that \code{DeadCodeEliminator} preserves the dynamic semantics:
the new code has the same execution behavior as the old code,
assuming the restriction about the absence of function definitions,
which is critical for the same reason explained above.
The theorem for statements is
\begin{bcode}
(defthm exec-statement-of-dead
  (implies (and (statement-nofunp stmt)
                (funenv-nofunp funenv))
           (soutcome-result-okeq
            (exec-statement
             (statement-dead stmt) cstate (funenv-dead funenv) limit)
            (exec-statement stmt cstate funenv limit))))
\end{bcode}
where:
\begin{itemize}[nosep]
\item
As in the previous theorem:
\code{stmt} is the old statement;
\code{(statement-dead~stmt)} is the new statement;
\code{statement-nofunp} is the critical restriction.
\item
\code{funenv-dead} extend the transformation to
(the function bodies in) function environments.
When a function is called during execution,
its body is retrieved from the function environment:
to apply induction hypotheses during the proof,
the function bodies in the old and new function environments
must be related by the transformation,
in the same way as the code being executed.
\item
\code{funenv-nofunp} extends the restriction of no function definitions to
(the function bodies in) function environments.
This is also needed to apply induction hypotheses during the proof,
because the code of called functions is retrieved from the function environment.
\item
\code{soutcome-result-okeq} is an equivalence relation on \code{soutcome-result}
that holds on \code{a} and \code{b} exactly when
either they are equal statement outcomes
or they are both error values.
This accommodates slight differences in
the details of the error values returned by the dynamic semantics.%
\footnote{These slight differences exist because the error values
contain some user-oriented information about the error causes,
e.g.\ the constructs that cause the errors.
But since this is unnecessary for verification,
the error values should probably be simplified,
only distinguishing between limit exhaustion and unsafe operations.
This should obviate the need for the equivalence relation.}
\item
The theorem says that,
assuming no function definitions in the old statement and function environment,
executing the old statement gives equivalent results to
executing the new statement on
the same computation state,
the transformed function environment,
and the same artificial limit.
\end{itemize}
The proof is not difficult,
but involves certain \code{:expand} and \code{:use} hints,
applied only to certain cases of the induction via computed hints,
because they slow down the proof if applied to all the cases.
Perhaps the \code{:use} hints could be avoided in some way,
but the \code{:expand} hints may be necessary
to defeat heuristics that prevent
the opening of certain recursive function calls.

The formulation of the theorem above does not distinguish between
errors due to limit exhaustion and errors due to unsafe operations.
In general, a transformation should not turn
terminating code into non-terminating code.
This can be proved by distinguishing between the two kinds of errors,
as done in the theorems for variable renaming, described next.

We proved that the variable renaming component of the \code{Disambiguator}
preserves both static and dynamic semantics.
Although this is intuitively obvious,
picturing the old and new code merely differing in variable names
but otherwise completely isomorphic,
it takes a bit of work to formulate and prove.

The theorem for the preservation of the static semantics for statements is
\begin{bcode}
(defthm check-safe-statement-when-renamevar
  (b* ((ren1 (statement-renamevar stmt-old stmt-new ren))
       (varmodes-old (check-safe-statement stmt-old (varset-old ren) funtab))
       (varmodes-new (check-safe-statement stmt-new (varset-new ren) funtab)))
    (implies (and (not (reserrp ren1))
                  (not (reserrp varmodes-old)))
             (and (not (reserrp varmodes-new))
                  (equal (vars+modes->vars varmodes-old)
                         (varset-old ren1))
                  (equal (vars+modes->vars varmodes-new)
                         (varset-new ren1))
                  (equal (vars+modes->modes varmodes-old)
                         (vars+modes->modes varmodes-new))))))
\end{bcode}
where:
\begin{itemize}[nosep]
\item
The theorem assumes that the new statement \code{stmt-new}
is a valid result of transforming the old statement \code{stmt-old},
given a renaming \code{ren},
which results in the possibly extended renaming \code{ren1}.
\item
The safety of the old/new statement is checked using the variable symbol table
consisting of the keys/values of the renaming \code{ren},
which are indeed the accessible variables,
given how the binary relation \code{statement-renamevar} is defined
(see \secref{sec:definitions}).
The same function symbol table \code{funtab} is used for both,
since functions are not renamed, only variables.%
\footnote{This is a motivation for decomposing \code{Disambiguator}
into the four independent components described in \secref{sec:definitions}.}
\item
The theorem assumes that the old statement is safe, and concludes that:
the new statement is safe too;
the updated variable symbol table after the old/new statement
consists of the keys/values of the updated renaming \code{ren1};
the termination modes of the new statement
are the same as the ones of the old statement.
\end{itemize}
The proof involves some preparatory lemmas,
as well as a custom induction scheme that takes into account
the recursive structure of
both the variable renaming functions like \code{statement-renamevar}
and the static semantic functions like \code{check-safe-statement}.

The theorem for the preservation of the dynamic semantics for statements is
\begin{bcode}
(defthm exec-statement-when-renamevar
  (b* ((ren1 (statement-renamevar stmt-old stmt-new ren)))
    (implies (and (not (reserrp ren1))
                  (cstate-renamevarp cstate-old cstate-new ren)
                  (funenv-renamevarp funenv-old funenv-new))
             (b* ((outcome-old
                   (exec-statement stmt-old cstate-old funenv-old limit))
                  (outcome-new
                   (exec-statement stmt-new cstate-new funenv-new limit)))
               (implies (and (not (reserr-nonlimitp outcome-old))
                             (not (reserr-nonlimitp outcome-new)))
                        (soutcome-result-renamevarp outcome-old
                                                    outcome-new
                                                    ren1))))))
\end{bcode}
where:
\begin{itemize}[nosep]
\item
The assumption involving \code{statement-renamevar}
is the same as in the previous theorem.
\item
\code{cstate-renamevarp} extends variable renaming
to computation states, which are built from the variables in the code.
The old and new computation states are renamed according to \code{ren}.
\item
\code{funenv-renamevarp} extends variable renaming
to function environments, from which the code of called functions is retrieved.
This does not depend on \code{ren},
because the variables in the body of every function in the environment
are renamed independently.
\item
\code{soutcome-renamevarp} (not directly used in the theorem)
extends variable renaming to statement outcomes \code{a} and \code{b}:
the computation states must be related by \code{cstate-renamevarp},
and the termination modes must be the same.
\item
\code{soutcome-result-renamevarp} extends variable renaming
to \code{soutcome-result}:
it holds on \code{a} and \code{b} exactly when
either they are both statement outcomes satisfying \code{soutcome-renamevarp}
or they are both error values.
\item
\code{reserr-nonlimitp} recognizes non-limit error values.
Thus, the theorem assumes that the execution of both old and new statement
results in either a limit error or a statement outcome.
\item
The theorem concludes that the two executions yield equivalent results.
\end{itemize}
The proof involves several preparatory lemmas,
a custom induction scheme that takes into account
the recursive structure of
both the variable renaming functions like \code{statement-renamevar}
and the execution functions like \code{exec-statement},
and several computed hints to apply different collections of common hints
to different cases of the induction.

It may seem strange that the above theorem assumes, instead of concluding,
that the execution of the new statement does not yield a non-limit error value:
compare the theorem for \code{DeadCodeEliminator}.
However, the preservation of the static semantics
by variable renaming described earlier,
and the general proof of static soundness in \secref{sec:static-soundness},
imply that the execution of the new statement
does not yield a non-limit error;
thus, it can be assumed in the theorem above,
making the proof slightly easier.

The dynamic semantics preservation theorems for \code{DeadCodeEliminator}
require transforming the function environment,
but not the computation state or the limit.
The dynamic semantics preservation theorems for variable renaming
require transforming the computation state and function environment,
but not the limit.
Theorems for more complex transformations
require transforming the limit as well,
because the number of execution steps can change.

\section{Related Work}
\label{sec:related}

We are not aware of any other work on Yul using ACL2.

Other formalizations of Yul exist, written in
generic math \cite{yul-math},
K \cite{yul-k},
Isabelle/HOL \cite{yul-isabelle},
Lean \cite{yul-lean,yul-lean-medium}, and
Dafny \cite{yul-dafny}.
As pointed out in \cite{yul-math},
none of these are peer-reviewed, except for \cite{yul-math} itself.
It does not appear that any of these formalizations include
a static semantics separate from the dynamic semantics,
and a static soundness proof relating the two.
On the other hand, \cite{yul-math} includes
both a small-step and a big-step dynamic semantics,
with a proof of equivalence.
An advantage of \cite{yul-math} compared to
the other formalizations, including ours,
is that it is written in a generic mathematical notation
that is more widely accessible than
the language and libraries of ACL2 and similar tools;
a disadvantage is that it is not machine-checked,
unlike the ones developed with ACL2 and similar tools.

The \code{README} in the K formalization of Yul \cite{yul-k} says that
its purpose is to perform translation validation of the Solidity compiler,
which is exactly the same goal as ours;
that \code{README} also indicates scripts to run tests.
It would be interesting to compare their work with ours,
but the lack of published papers and detailed documentation
demands an examination of their code
and prerequisite knowledge of K.
The stated purpose of \cite{yul-math}
is to provide a widely accessible precise formalization of Yul.
The purpose of \cite{yul-isabelle,yul-lean,yul-dafny}
appears to be mainly the formal verification of Yul code,
but not specifically Yul transformations.

\section{Future Work}
\label{sec:future}

Although our formalization of the Yul core is essentially complete,
it hard-wires some aspects of the EVM dialect,
which should better be kept more separate
via a more explicit parameterization of the core formalization over the dialect.
This may take some effort because, despite the multi-dialect aspiration,
currently the Yul core and the EVM dialect are not crisply delineated,
and manifest very much like one integrated language;
for instance, different dialects are supposed to have different type systems,
but the Yul syntax does not provide a way to specify types.%
\footnote{The Yul documentation includes
an older grammar with syntax for type identifiers,
and a newer grammar without such syntax.
The Yul team from the Ethereum Foundation
told us that the latter supersedes the former.}

More importantly than having a cleaner separation between core and dialects,
the EVM dialect should be formalized completely,
since it is currently the only Yul dialect in practical use.
This is a laborious task because it involves
modeling a large portion of the functionality of the EVM \cite{yellow-paper},
although it does not require a full formalization of the EVM itself.

Our work on transformations has barely scratched the surface.
The Solidity compiler includes tens of transformations,
some of which are rather complex
and involve EVM-dialect-specific features.
Formalizing and verifying all of them is a substantial task,
but it can be approached piecewise:
each transformation can be formalized and verified mostly on its own;
dependencies among transformations can be handled by formalizing
the kinds of restrictions on Yul code exemplified in \secref{sec:restrictions}.

To verify the transformations performed by the Solidity compiler,
the theorem generator depicted in \figref{fig:approach} must be developed.
Once all transformations have been formalized and verified,
the generated proofs can be composed into a proof that
the entire sequence of Yul transformations was correctly applied.

\section{Conclusion}
\label{sec:conclusion}

Our formalization of Yul, like formalizations developed by others,
does not contain any particularly innovative ideas,
partly because Yul is a relatively simple language.
However, a precise, machine-checked formalization of Yul is clearly valuable.
Furthermore, it is possible that
extending the formalization to the full EVM dialect
may uncover more interesting formalization issues.

The work on verifying Yul transformations is more original,
though there is some similar work (see \secref{sec:related}).
An interesting finding of our work was that
verifying the correctness of even fairly simple transformations,
such as those described in \secref{sec:transformations},
whose correctness intuitively appears obvious,
required more work than expected.
The proofs are not difficult, but a bit laborious.


\section*{Acknowledgements}

We thank the Ethereum Foundation for supporting this work.


\bibliographystyle{eptcs}
\bibliography{paper}


\end{document}